\documentclass[conference]{IEEEtran}

\usepackage[cmex10]{amsmath}
\usepackage{color,graphicx}
\usepackage{citesort}

\newtheorem{theorem}{Theorem}

\newtheorem{remark}{Remark}
\IEEEoverridecommandlockouts

\begin{document}
\title{On the Sum Capacity of the Gaussian X Channel in the Mixed Interference Regime}
\author{\IEEEauthorblockN{Praneeth~Kumar~V\thanks{This work was done at the Department of Electrical Engineering, Indian Institute of Technology Madras, Chennai, India.}}
\IEEEauthorblockA{ISRO Satellite Center\\ 
Bangalore 560017, India.\\
Email: praneethkumar7@gmail.com}
\and
\IEEEauthorblockN{Srikrishna Bhashyam}
\IEEEauthorblockA{Department of Electrical Engineering\\
Indian Institute of Technology Madras, Chennai 600036.\\
Email: skrishna@ee.iitm.ac.in}}
\maketitle

\begin{abstract}
In this paper, we analyze the Gaussian X channel in the mixed interference regime. In this regime, multiple access transmission to one of the receivers is shown to be close to optimal in terms of sum rate. Three upper bounds are derived for the sum capacity in the mixed interference regime, and the subregions where each of these bounds dominate the others are identified. The genie-aided sum capacity upper bounds derived also show that the gap between sum capacity and the sum rate of the multiple access transmission scheme is small for a significant part of the mixed interference region. For any $\delta > 0$, the region where multiple access transmission to one of the receivers is within $\delta$ from sum capacity is determined. 
\end{abstract}


\IEEEpeerreviewmaketitle

\section{Introduction and problem statement}
\label{sec:intro}
\begin{figure}[!ht]
\centering
\def\svgwidth{0.85\columnwidth}
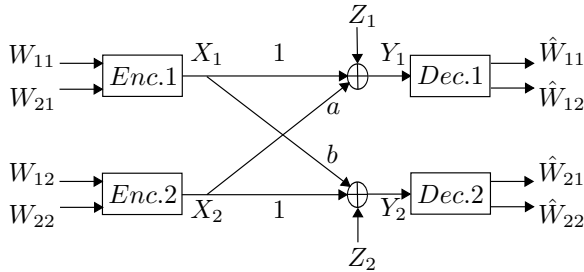
\vspace*{-1mm}
\caption{Two user scalar Gaussian X channel}
\label{fig:macgap:gxc}
\end{figure}

The Gaussian X channel consists of two transmitters and two receivers where each transmitter has an independent message for each receiver (see Fig. \ref{fig:macgap:gxc}).
A scalar Gaussian X channel in standard form is described by the following equations: 
\begin{IEEEeqnarray}{rCl}
\label{eqn:macgap:gxc1}
Y_1&=&X_1 +aX_2 + Z_1, \\*
\label{eqn:macgap:gxc2}
Y_2&=&bX_1 +X_2 + Z_2, 
\end{IEEEeqnarray}
where $Z_1, Z_2\sim \mathcal{N}(0,1)$ and i.i.d in time. Transmitter $i$ has an average power constraint $P_i$ i.e., $\frac{1}{n} \sum_{k = 1}^{n} E\left[X_{ik}^{2}\right] \leq P_i$. The interference channel (IC) can be obtained as a special case of the X channel by using only messages $W_{11}$ and $W_{22}$. 


The sum capacity of the Gaussian Z channel is known \cite{ChoMotGar07}. The sum capacity of the Gaussian IC is known for the strong \cite{Car75}, mixed \cite{MotKha09}, and noisy interference regimes \cite{AnnVee09,MotKha09,ShaKraChe09}. The exact capacity region or even the sum capacity of the Gaussian X channel are not fully known. Capacity approximations, degrees of freedom results, and sum capacity for some channel conditions are available in \cite{CadJaf09,MadMotKha08,HuaCadJaf12,NieMad13}. In \cite{CadJaf09,MadMotKha08}, degrees of freedom results were derived and the interference alignment technique was proposed. In \cite{HuaCadJaf12}, the sum capacity of the Gaussian X channel was determined in the noisy interference regime. This result is an extension of a similar result for the Gaussian IC in \cite{AnnVee09}. The {\em noisy interference} regime corresponds to the set of channel conditions (conditions on $a$ and $b$) under which treating interference from the cross links as noise is sum capacity optimal for the Gaussian IC \cite{AnnVee09}. For the same regime, treating interference as noise is also sum capacity optimal for the Gaussian XC. Recently, in \cite{NieMad13}, 
the capacity region of the 
Gaussian X channel was obtained to within a constant gap of 4 bits. The focus in \cite{NieMad13} was on channel conditions where interference alignment is necessary. Channel realizations where all channel gains are small ($< 1$) are ignored since their capacity is bounded by a constant less than 4. To the best of our knowledge, the only known sum capacity bounds for the Gaussian X channel are the bounds in \cite{CadJaf09,MadMotKha08,HuaCadJaf12,NieMad13}.

In this paper, we focus on the sum capacity of the Gaussian X channel in the mixed interference regime, i.e., the regions ${\cal R}_1 = \{ (a, b): a^2 \ge 1, b^2 \le 1\}$ and ${\cal R}_2 = \{ (a, b): a^2 \le 1, b^2 \ge 1\}$.
The sum capacity and capacity region of the Gaussian IC  in the mixed interference regime  have been studied in \cite{MotKha09}.  The sum capacity is shown to be achieved by the strategy that achieves sum capacity for the one-sided weak Gaussian IC or the one-sided strong Gaussian IC. The analogous one-sided channel for the X channel is the Z channel \cite{ChoMotGar07} obtained by setting either $a = 0$ or $b = 0$. In \cite{ChoMotGar07}, it was shown that, for $b=0$ and $a^2 \ge 1$, the sum capacity is equal to the multiple access channel (MAC) capacity at receiver 1. Similarly, for $a=0$ and $b^2 \ge 1$, the sum capacity is equal to the multiple access channel (MAC) capacity at receiver 2. In this paper, we ask the following question:  Is MAC at one of the receivers sum rate optimal or approximately sum rate optimal for some subregion of the mixed interference region of the Gaussian X channel? Note that the sum rate of the MAC at one of the receivers grows unbounded as $a$ or $b$ increases. Therefore, even 
an approximate result with small or finite rate gap is useful.

The main results in this paper are as follows. 
\begin{enumerate}
\item A genie-aided sum capacity upper bound, $C_U^A$, valid in the region $\{ (a, b): a^2 > (P_1 + 1)^2, b^2 \le 1\} \subset {\cal R}_1$ is derived for the Gaussian X channel. An expression for the gap between this upper bound and the MAC sum rate at receiver 1 is derived. The gap goes to zero as $a^2 \rightarrow \infty$. 
\item Another genie-aided sum capacity upper bound, $C_U^B$, valid in the region $\{ (a, b): a^2 \ge 1, b^2 < 1/(a^2P_2 + 1)^2\} \subset {\cal R}_1$ is derived for the Gaussian X channel. Again, an expression for the gap between this upper bound and the MAC sum rate at receiver 1 is derived.
\item A simple sum capacity upper bound, $C_U^C$, valid in the whole region ${\cal R}_1$, is then compared with the above two genie-aided bounds. The sub-regions where each upper bound provides the best bound are determined.
\item Using these bounds, the regime of possible $(a,b)$ where the gap between the upper and lower bounds is less than an arbitrary $\delta > 0$ is obtained (See Remark \ref{rem:rdelta}). We denote this regime ${\cal R}_{\delta}$ (note ${\cal R}_{\delta} \subset {\cal R}_1$). Thus, MAC at receiver 1 is $\delta$ close to optimal in terms of sum rate in this subregion.
\item Similar results can be obtained for subregions of ${\cal R}_2$ by interchanging the roles of $a$ and $b$, and $P_1$ and $P_2$. In this subregion, MAC at receiver 2 is $\delta$ close to optimal in terms of sum rate.
\end{enumerate}
\section{Sum capacity bounds in Region ${\cal R}_1$}
In this section, we derive genie-aided upper bounds for sum capacity of the Gaussian X channel in ${\cal R}_1$. We compare these bounds with the sum rate achieved by MAC at receiver 1.

\subsection{MAC strategy at receiver 1 and its sum rate}
The strategy corresponding to {\em MAC at receiver 1} is the following. Messages to receiver 2, $W_{21}, W_{22}$, are not transmitted, and messages $W_{11},W_{12}$ are transmitted using the optimal Gaussian MAC coding strategy to receiver 1. The sum rate achieved by this strategy is given by:
\begin{equation}
R_{MAC,1} = \frac{1}{2}\log_2\left( 1 + P_1 + a^2P_2\right).
\label{MAC1sumrate}
\end{equation}

\subsection{Genie-aided sum capacity upper bound $C_U^A$}
\label{sec:mac1}
\begin{theorem}
\label{thm1}
(a) For a Gaussian X channel with $b^2\leq 1$ and $a^2 > (P_1+1)^2$, the sum capacity of X channel $C_{sum}$ is bounded as follows:
\begin{equation}
C_{sum} \le R_{MAC,1} + \frac{1}{2}\log_2 \left( \frac{1 - \frac{P_1+1}{a^2}}{1 - \frac{(P_1+1)^2}{a^2}}\right) =: C_U^A. \label{thm:upperbound1}
\end{equation}
(b) For a Gaussian X channel with $b^2\leq 1$ and $a^2 > \frac{(P_1+1)[(P_1+1)2^{2\delta} - 1]}{2^{2\delta} - 1}$, we have $C_{sum} - R_{MAC,1} < \delta.$
\end{theorem}
\begin{IEEEproof}
(a) Consider the genie-aided X channel with side information $S_1$  given to the receiver 1. 
Let the side information be $S_1=X_2+\eta W$ where $W\sim \mathcal{N}(0,1)$, and $W$  is correlated with $Z_1$ with correlation coefficient $\rho$. We will find an upper bound on the sum capacity of the genie-aided channel which in turn is an upper bound to the original X channel.

Intuitively, we will argue all the messages decodable in the original channel are decodable by the genie enhanced receiver 1. $W_{11}$, $W_{12}$ can be decoded reliably by the enhanced receiver 1 as they can be decoded in the original channel. $W_{22}$ can be decoded from $S_1$ at the receiver 1 for $\eta^2 \le 1$ because receiver 2 in the original channel can decode it from $Y_2$. Then, using the decoded $W_{12}$, $W_{22}$, the enhanced receiver 1 decodes $W_{21}$ by cancelling the effect of $X_2$ to get $X_1+Z_1$, which is a less noisy signal than $Y_2$ for $b^2 \le 1$. Finally, we choose $\rho$ to reduce the gap between the sum capacity upper bound and the sum rate of the MAC at receiver 1.
\allowdisplaybreaks{\begin{IEEEeqnarray}{rCl}
\IEEEeqnarraymulticol{3}{l}{
n(R_{11}+R_{12}+R_{21}+R_{22})=H(W_{11},W_{12},W_{21},W_{22})
}\IEEEnonumber\\ \qquad \quad
&=&I(W_{11},W_{12},W_{21},W_{22};Y_1^n,S_1^n) \IEEEnonumber \\ &&+\:H(W_{11},W_{12},W_{21},W_{22}|Y_1^n,S_1^n)
\end{IEEEeqnarray}}
First, expand and bound the term $H(W_{11},W_{12},W_{21},W_{22}|Y_1^n,S_1^n)$ as follows.
\allowdisplaybreaks{\begin{IEEEeqnarray}{rCl}
\IEEEeqnarraymulticol{3}{l}{
H(W_{11},W_{12},W_{21},W_{22}|Y_1^n,S_1^n)}\IEEEnonumber \\ \quad&=&H(W_{11}|Y_1^n,S_1^n)+H(W_{12}|Y_1^n,S_1^n,W_{11})\IEEEnonumber\\
&&+\:H(W_{22}|Y_1^n,S_1^n,W_{11},W_{12})\IEEEnonumber\\
&&+\:H(W_{21}|Y_1^n,S_1^n,W_{11},W_{12},W_{22})\\
&\stackrel{(a)}{\leq}&H(W_{11}|Y_1^n,S_1^n)+H(W_{12}|Y_1^n,S_1^n)+H(W_{22}|S_1^n)\IEEEnonumber\\
&&+\:H(W_{21}|Y_1^n,W_{12},W_{22}),
\label{eqn:macgap:gxc10}
\end{IEEEeqnarray}}
where step $(a)$ follows from the fact that conditioning reduces the entropy. Since $W_{11}$ and $W_{12}$ have to be decodable at receiver 1, using Fano's inequality, the first two terms in (\ref{eqn:macgap:gxc10}) can be bounded by $n(\epsilon_{1n}+\epsilon_{2n})$, where $\epsilon_{1n}$ and $\epsilon_{2n}$ tend to 0 as $n \rightarrow \infty$. The next two terms can be bounded under the conditions assumed for $a^2$ and $b^2$ in the theorem statement.

Consider the fourth term $H(W_{21}|Y_1^n,W_{12},W_{22})$ in (\ref{eqn:macgap:gxc10}).
\allowdisplaybreaks{\begin{IEEEeqnarray}{rCl}
\IEEEeqnarraymulticol{3}{l}{
H(W_{21}|Y_1^n,W_{12},W_{22})}\IEEEnonumber \\ 
\quad&=& H(W_{21}|X_1^n +aX_2^n + Z_2^n,W_{12},W_{22}) \IEEEnonumber\\
&\stackrel{(b)}{=}& H(W_{21}|X_1^n + Z_2^n,W_{12},W_{22})\IEEEnonumber\\
&\stackrel{(c)}{\leq}&H(W_{21}|X_1^n + \frac{Z_2^n}{b},W_{12},W_{22})\IEEEnonumber\\
& = & H(W_{21}|bX_1^n + Z_2^n,W_{12},W_{22})\IEEEnonumber\\
&\stackrel{(b)}{=} & H(W_{21}|bX_1^n + X_2^n + Z_2^n,W_{12},W_{22})\IEEEnonumber\\
& = & H(W_{21}|Y_2^n,W_{12},W_{22})\IEEEnonumber\\
&\stackrel{(d)}{\leq}& H(W_{21}|Y_2^n) \stackrel{(e)}{\leq} n\epsilon_{4n}, \label{eqn:eps4}
\end{IEEEeqnarray}}
where $(b)$ holds since $X_2^n$ is a deterministic function of $W_{12},W_{22}$, $(c)$ holds since $X_1^n + Z_2^n$ is a less noisy version than $X_1^n + \frac{Z_2^n}{b}$ for $b^2\leq1$, $(d)$ holds because conditioning reduces entropy, and $(e)$ holds by Fano's inequality and $\epsilon_{4n}$ tends to 0 as $n \rightarrow \infty$.

Now consider the third term $H(W_{22}|S_1^n)$ in (\ref{eqn:macgap:gxc10}).
\allowdisplaybreaks{\begin{IEEEeqnarray}{rCl}
H(W_{22}|S_1^n)&=& H(W_{22}|X_2^n +\eta W^n)\IEEEnonumber\\
&\stackrel{(f)}{\leq}& H(W_{22}|X_2^n +Z_2^n)\IEEEnonumber\\
&\stackrel{(g)}{=}& H(W_{22}|X_2^n +Z_2^n,W_{11},W_{21})\IEEEnonumber\\
&\stackrel{(h)}{=}& H(W_{22}|bX_1^n +X_2^n + Z_2^n,W_{11},W_{21})\IEEEnonumber\\
&\stackrel{(i)}{\le}& H(W_{22}|Y_2^n) \stackrel{(j)}{\leq} n\epsilon_{3n}, \label{eqn:eps3}
\end{IEEEeqnarray}}
where $(f)$ holds since $X_2^n +\eta W^n$ is a less noisy version of $X_2^n +Z_2^n$ for $\eta^2\leq1$, $(g)$ holds since $W_{11}$ and $W_{21}$ are independent of the other variables $W_{22}$, $X_2^n$ and $Z_2^n$, $(h)$ holds since $X_1^n$ is a function of $W_{11}, W_{21}$, $(i)$ holds because conditioning reduces entropy, and $(j)$ holds by Fano's inequality and $\epsilon_{3n}$ tends to 0 as $n \rightarrow \infty$. Therefore, we have $H(W_{11},W_{12},W_{21},W_{22}|Y_1^n,S_1^n)\leq n(\epsilon_{1n}+\epsilon_{2n}+\epsilon_{3n}+\epsilon_{4n}) =: n\epsilon$, and
\allowdisplaybreaks{\begin{IEEEeqnarray}{rCl}
\IEEEeqnarraymulticol{3}{l}{
n(R_{11}+R_{12}+R_{21}+R_{22}-\epsilon)}\IEEEnonumber \\ \quad &\leq&I(W_{11},W_{12},W_{21},W_{22};Y_1^n,S_1^n) \label{eqnsum1}\\
&=&h(Y_1^n,S_1^n)-h(Y_1^n,S_1^n|W_{11},W_{12},W_{21},W_{22})\IEEEnonumber\\
&=&h(Y_1^n,S_1^n)-nh(Z_1, \eta W)\IEEEnonumber\\
&\stackrel{(j)}{\leq}&nh(Y_{1G},S_{1G})- nh(Z_1, \eta W) \IEEEnonumber\\
&=&nI(X_{1G},X_{2G};Y_{1G},S_{1G}) \IEEEnonumber\\
&=&nI(X_{1G},X_{2G};Y_{1G})+ nI(X_{1G},X_{2G};S_{1G}|Y_{1G})\IEEEnonumber\\
&=&nR_{MAC,1}+ nI(X_{1G},X_{2G};S_{1G}|Y_{1G}) \label{eqnsum2},
\end{IEEEeqnarray}}
where step (j) is because the Gaussian distribution maximizes the joint entropy term. Now consider $I(X_{1G},X_{2G};S_{1G}|Y_{1G}) 
= I(X_{2G};S_{1G}|Y_{1G})+I(X_{1G};S_{1G}|Y_{1G},X_{2G})$. 
The first term can be made zero if \cite[Lemma 8]{AnnVee09} 
\begin{equation}
\eta \rho=\frac{P_1+1}{a}. \label{eqn:etarho}
\end{equation}
The second term $I(X_{1G};S_{1G}|Y_{1G},X_{2G})$ cannot be made zero and is, therefore, the gap between the sum capacity upper bound and sum rate achieved by MAC at receiver 1.
\allowdisplaybreaks{\begin{IEEEeqnarray}{rCl}
\IEEEeqnarraymulticol{3}{l}{
I(X_{1G};S_{1G}|Y_{1G},X_{2G})}\IEEEnonumber \\ \quad&=&I(X_{1G};X_{2G}+\eta W|X_{1G}+aX_{2G}+Z_1,X_{2G})\\
&=&h(X_{2G}+\eta W|X_{1G}+aX_{2G}+Z_1,X_{2G})\\
&&-\:h(X_{2G}+\eta W|X_{1G}+aX_{2G}+Z_1,X_{1G},X_{2G})\quad \\
&=&h(\eta W|X_{1G}+Z_1)-h(\eta W|Z_1)\\
&=&\frac{1}{2}\log_2\left(\frac{1-\frac{\rho^2}{P_1+1}}{1-\rho^2}\right) \text{bits} \label{eqn:gaprho}
\end{IEEEeqnarray}}
Note that we can choose any $\rho^2 \le 1$ as long as (\ref{eqn:etarho}) and $\eta^2 \le 1$ are also satisfied, i.e., we need to satisfy $1 \ge \rho^2 \ge \frac{(P_1+1)^2}{a^2}$. Since (\ref{eqn:gaprho}) is increasing in $\rho$, to get the smallest upper bound, we choose $\rho^2 = \frac{(P_1+1)^2}{a^2}$. This gives the upper bound (\ref{thm:upperbound1}) in the statement (a) of the theorem. 

(b) The gap is bounded by $\delta$ if $\frac{1}{2}\log_2 \left( \frac{1 - \frac{P_1+1}{a^2}}{1 - \frac{(P_1+1)^2}{a^2}}\right) < \delta$. Simplifying this inequality results in statement (b) of the theorem.\end{IEEEproof}
\begin{remark}
\label{rem1}
The gap between the sum capacity upper bound and the sum rate of MAC at receiver 1 reduces with increasing $a^2$ and goes to zero as $a^2 \rightarrow \infty$ (Some numerical examples are shown in Fig. \ref{fig:sumratea2} of Section \ref{sec:results}). To see this, rewrite the gap in the form $\ln \left(\frac{1-\frac{x}{v+1}}{1-x}\right)$, where $v = P_1$ and $ 0 \leq x = \frac{(P_1 + 1)^2}{a^2} < 1 $. Using Taylor's series, $\ln\left(1-x\right)=-\sum_{n=1}^\infty \frac{x^n}{n}, \mbox{~for~} |x|<1~$, and we can bound as follows 
\begin{IEEEeqnarray}{rCl}
 \left(\frac{v}{1+v}\right)\ln\left(\frac{1}{1-x}\right) &\leq \ln\left(\frac{1-\frac{x}{v+1}}{1-x}\right) &\leq \ln\left(\frac{1}{1-x}\right).~~
\end{IEEEeqnarray}
for $ 0 \leq x < 1 $.
As $x \rightarrow 0$, $\ln\left(\frac{1}{1-x}\right)$ goes to zero linearly in $x$. Therefore, the gap between the upper and lower bounds approches zero as $a^2 \rightarrow \infty$. 

\end{remark}

\begin{remark}
The MAC at receiver 1 is $\delta$ close to optimal in terms of sum capacity for most of the mixed interference regime ${\cal R}_1$ except for a finite subregion. 
\end{remark}

\begin{remark}
Unlike the approximate sum capacity result in this paper, an exact sum capacity result for a subregion of the mixed interference regime is claimed in  \cite[Sec. 6, Thm. 4]{SriBha12}. However, there is an error in step (iii) of the proof of Thm. 4 in \cite{SriBha12} which allows the gap between the upper bound and the MAC strategy sum rate to go to 0.
\end{remark}

\subsection{Genie-aided sum capacity upper bound $C_U^B$}
\label{sec:mac2}
\begin{theorem}
\label{thm2}
(a) For a Gaussian X channel with $a^2\geq 1$ and $b^2 < \frac{1}{(a^2P_2+1)^2}$, the sum capacity of X channel $C_{sum}$ is bounded as follows:
\begin{equation}
C_{sum} \le R_{MAC,1} + \frac{1}{2}\log_2 \left( \frac{1 - b^2(a^2P_2 + 1)}{1 - b^2(a^2P_2 + 1)^2}\right) =: C_U^B. \label{thm:upperbound2}
\end{equation}
(b) For a Gaussian X channel with $a^2\geq 1$ and $b^2 < \frac{(2^{2\delta} - 1)}{((a^2P_2 + 1)2^{2\delta} - 1)(a^2P_2 + 1)}$, we have $C_{sum} - R_{MAC,1} < \delta.$
\end{theorem}
\begin{IEEEproof}
(a) The proof technique is similar to the proof of Theorem \ref{thm1} (a) except that a different genie is chosen. Consider the genie-aided X channel with side information $S_1=bX_1+\eta W$, where $W\sim \mathcal{N}(0,1)$ is correlated with $Z_1$ and $\rho$ is the correlation coefficient. A similar intuitive argument as in Theorem \ref{thm1} can also be made here. Here, $W_{21}$ can be decoded from $S_1$ at the receiver 1 for $\eta^2 \le 1$ because receiver 2 in the original channel can decode it from $Y_2$. Then, using the decoded $W_{11}$, $W_{21}$, the enhanced receiver 1 decodes $W_{22}$ by cancelling the effect of $X_1$ to get $X_2+\frac{Z_1}{a}$, which is a less noisy signal than $Y_2$ for $a^2 \ge 1$. 

Expand and bound the term $H(W_{11},W_{12},W_{21},W_{22}|Y_1^n,S_1^n)$ as follows.
\allowdisplaybreaks{\begin{IEEEeqnarray}{rCl}
\IEEEeqnarraymulticol{3}{l}{
H(W_{11},W_{12},W_{21},W_{22}|Y_1^n,S_1^n)}\IEEEnonumber \\ \quad
&{\leq}&H(W_{11}|Y_1^n,S_1^n)+H(W_{12}|Y_1^n,S_1^n)+H(W_{21}|S_1^n)\IEEEnonumber\\
&&+\:H(W_{22}|Y_1^n,W_{11},W_{21}), \label{eqn:terms}
\end{IEEEeqnarray}}
Since $W_{11}$ and $W_{12}$ have to be decodable at receiver 1, using Fano's inequality, the first two terms in (\ref{eqn:terms}) can be bounded by $n(\epsilon_{1n}+\epsilon_{2n})$, where $\epsilon_{1n}$ and $\epsilon_{2n}$ tend to 0 as $n \rightarrow \infty$. 

As done in (\ref{eqn:eps4}), the fourth term $H(W_{22}|Y_1^n,W_{11},W_{21})$ in (\ref{eqn:terms}) can be bounded as $H(W_{22}|Y_1^n,W_{11},W_{21}) \le H(W_{22}|Y_2^n) \le n\epsilon_{4n}$ for $a^2\geq1$ since $X_2^n + \frac{Z_2^n}{a}$ is a less noisy version than $X_2^n + Z_2^n$. As done in (\ref{eqn:eps3}), the third term $H(W_{21}|S_1^n)$ in (\ref{eqn:terms}) can be bounded as $H(W_{21}|S_1^n) \le H(W_{21}|Y_2^n) \le n\epsilon_{3n}$ for $\eta^2\leq1$. Therefore, we have $H(W_{11},W_{12},W_{21},W_{22}|Y_1^n,S_1^n)\leq n(\epsilon_{1n}+\epsilon_{2n}+\epsilon_{3n}+\epsilon_{4n}) =: n\epsilon$. As done in equations (\ref{eqnsum1}) to (\ref{eqnsum2}), it can be shown that
\allowdisplaybreaks{\begin{IEEEeqnarray}{rCl}
\IEEEeqnarraymulticol{3}{l}{
n(R_{11}+R_{12}+R_{21}+R_{22}-\epsilon)}\IEEEnonumber \\ \qquad &\le& nR_{MAC,1}+ nI(X_{1G},X_{2G};S_{1G}|Y_{1G}). \IEEEnonumber
\end{IEEEeqnarray}}
Now consider  $I(X_{1G},X_{2G};S_{1G}|Y_{1G})
= I(X_{1G};S_{1G}|Y_{1G})+I(X_{2G};S_{1G}|Y_{1G},X_{1G})$.
The first term can be made zero if \cite[Lemma 8]{AnnVee09}
\begin{equation}
\eta \rho=b(a^2P_2+1). \label{eqn:etarho2}
\end{equation}
The second term $I(X_{2G};S_{1G}|Y_{1G},X_{1G})$ cannot be made zero and is, therefore, the gap between the sum capacity upper bound and sum rate achieved by MAC at receiver 1. It can be shown that
\begin{equation} 
I(X_{2G};S_{1G}|Y_{1G},X_{1G}) = \frac{1}{2}\log_2\left(\frac{1-\frac{\rho^2}{a^2P_2+1}}{1-\rho^2}\right) \text{bits}. \label{eqn:gaprho2}
\end{equation}
Note that we can choose any $\rho^2 \le 1$ as long as (\ref{eqn:etarho2}) and $\eta^2 \le 1$ are also satisfied, i.e., we need to satisfy $1 \ge \rho^2 \ge b^2(a^2P_2+1)^2$. Since (\ref{eqn:gaprho2}) is increasing in $\rho$, to get the smallest upper bound, we choose $\rho^2 = b^2(a^2P_2+1)^2$. This gives the upper bound (\ref{thm:upperbound2}) in the statement (a) of the theorem. 

(b) The gap is bounded by $\delta$ if $\frac{1}{2}\log_2 \left( \frac{1 - b^2(a^2P_2 + 1)}{1 - b^2(a^2P_2 + 1)^2}\right) < \delta$. Simplifying this inequality results in statement (b) of the theorem.
\end{IEEEproof}

\begin{remark}
The gap between the sum capacity upper bound and the sum rate of MAC at receiver 1 reduces with decreasing $b^2$ and goes to zero as $b^2 \rightarrow 0$ (based on the argument in Remark \ref{rem1}). 
\end{remark} 

\subsection{Sum capacity upper bound $C_U^C$}
\label{sec:mac3}
\begin{theorem}
\label{thm3}
(a) For a Gaussian X channel with $a^2\geq 1$, the sum capacity of X channel $C_{sum}$ is bounded as follows:
\begin{equation}
C_{sum} \le R_{MAC,1} + \frac{1}{2}\log_2 \left( 1 + b^2P_1\right) =: C_U^C. \label{thm:upperbound3}
\end{equation}
(b) For a Gaussian X channel with $a^2\geq 1$ and $b^2 < \frac{(2^{2\delta} - 1)}{P_1}$, we have $C_{sum} - R_{MAC,1} < \delta.$
\end{theorem}
\begin{IEEEproof}
(a) The sum of the rates of the messages $W_{11}$, $W_{12}$, and $W_{22}$ is upper bounded by the sum capacity of the Z channel obtained by setting $b = 0$ \cite{JafSha08}. The sum capacity of the Z channel for $a^2 \ge 1$ is $R_{MAC,1}$ \cite{ChoMotGar07}. Thus, we have
\begin{equation}
R_{11} + R_{12} + R_{22} \le R_{MAC,1}.
\label{eqn:bound3-1}
\end{equation}
It can also be shown easily that
\begin{equation}
R_{21} \le \frac{1}{2}\log_2 \left( 1 + b^2P_1\right).
\label{eqn:bound3-2}
\end{equation}
Combining (\ref{eqn:bound3-1}) and (\ref{eqn:bound3-2}), we get the bound on sum capacity in statement (a) of the theorem.

(b) The gap is bounded by $\delta$ if $\frac{1}{2}\log_2 \left( 1 + b^2P_1\right) < \delta$. Simplifying this inequality results in statement (b) of the theorem.
\end{IEEEproof}

\begin{remark}
\label{rem:rdelta}
Combining Theorems \ref{thm1}(b), \ref{thm2}(b) and \ref{thm3}(b), we get ${\cal R}_{\delta} = \{ (a,b): a^2 > \frac{(P_1+1)[(P_1+1)2^{2\delta} - 1]}{2^{2\delta} - 1}, b^2 \le 1\} \bigcup \{ (a,b): a^2 \ge 1, b^2 < \frac{(2^{2\delta} - 1)}{((a^2P_2 + 1)2^{2\delta} - 1)(a^2P_2 + 1)}\} \bigcup \{ (a,b): a^2 \ge 1, b^2 < \frac{(2^{2\delta} - 1)}{P_1}\}$.
\end{remark}

\subsection{Comparison of the upper bounds}
\label{subsec:comparison}
For a given $(a, b)$, the minimum of the upper bounds (i.e., $C_U^A$, $C_U^B$, and $C_U^C$ ) that are defined at $(a,b)$  is also an upper bound. Here, we identify the subregions where each bound is minimum. Using the difference between the expressions for the bounds, we can easily show that:

(1) $C_U^A < C_U^C$ when $a^2 > (P_1+1)^2 + \frac{(P_1+1)}{b^2}$.

(2) $C_U^B < C_U^C$ when $b^2 < \frac{1}{(a^2P_2+1)}\left[\frac{1}{(a^2P_2+1)}  - \frac{a^2P_2}{P_1}\right]$.

(3)  $C_U^B < C_U^A$ when $b^2 < \frac{1}{(a^2P_2+1)}\left[\frac{1}{1  + \frac{a^2P_2(a^2 - (P_1+1))}{P_1(P_1 + 1)}}\right]$.

A graphical illustration of the regions where each bound is minimum is shown for a numerical example in Section \ref{sec:results}. $C_U^B$ is minimum only when $P_2$ is small compared to $P_1$.

\section{Sum capacity bounds in Region ${\cal R}_2$}
In region ${\cal R}_2$, we have $a^2 \le 1$, $b^2 \ge 1$. By the symmetry of the model, results similar to theorems 1 and 2 can be obtained by interchanging $a$ and $b$, and interchanging $P_1$ and $P_2$. Here, we write (without proof) only the upper bounds on the sum capacity similar to Theorems \ref{thm1}(a), \ref{thm2}(a), and \ref{thm3}(a). 

\begin{theorem}
\label{thm4}
For a Gaussian X channel with $a^2\leq 1$ and $b^2 > (P_2+1)^2$, the sum capacity of X channel $C_{sum}$ is bounded as follows:
\begin{equation}
C_{sum} \le R_{MAC,2} + \frac{1}{2}\log_2 \left( \frac{1 - \frac{P_2+1}{b^2}}{1 - \frac{(P_2+1)^2}{b^2}}\right), \label{thm:upperbound4}
\end{equation}
where $R_{MAC,2} = \frac{1}{2}\log_2(1 + b^2P_1 + P_2)$ is the sum rate achieved by MAC at receiver 2.
\end{theorem}

\begin{theorem}
\label{thm5}
For a Gaussian X channel with $b^2\geq 1$ and $a^2 < \frac{1}{(b^2P_1+1)^2}$, the sum capacity of X channel $C_{sum}$ is bounded as follows:
\begin{equation}
C_{sum} \le R_{MAC,2} + \frac{1}{2}\log_2 \left( \frac{1 - a^2(b^2P_1 + 1)}{1 - a^2(b^2P_1 + 1)^2}\right). \label{thm:upperbound5}
\end{equation}
\end{theorem}

\begin{theorem}
\label{thm6}
For a Gaussian X channel with $b^2\geq 1$, the sum capacity of X channel $C_{sum}$ is bounded as follows:
\begin{equation}
C_{sum} \le R_{MAC,2} + \frac{1}{2}\log_2 \left( 1 + a^2P_2\right). \label{thm:upperbound6}
\end{equation}
\end{theorem}


\section{Numerical Results}
 \label{sec:results}
 \begin{figure}[htbp]
\centering
\includegraphics[width=0.95\columnwidth]{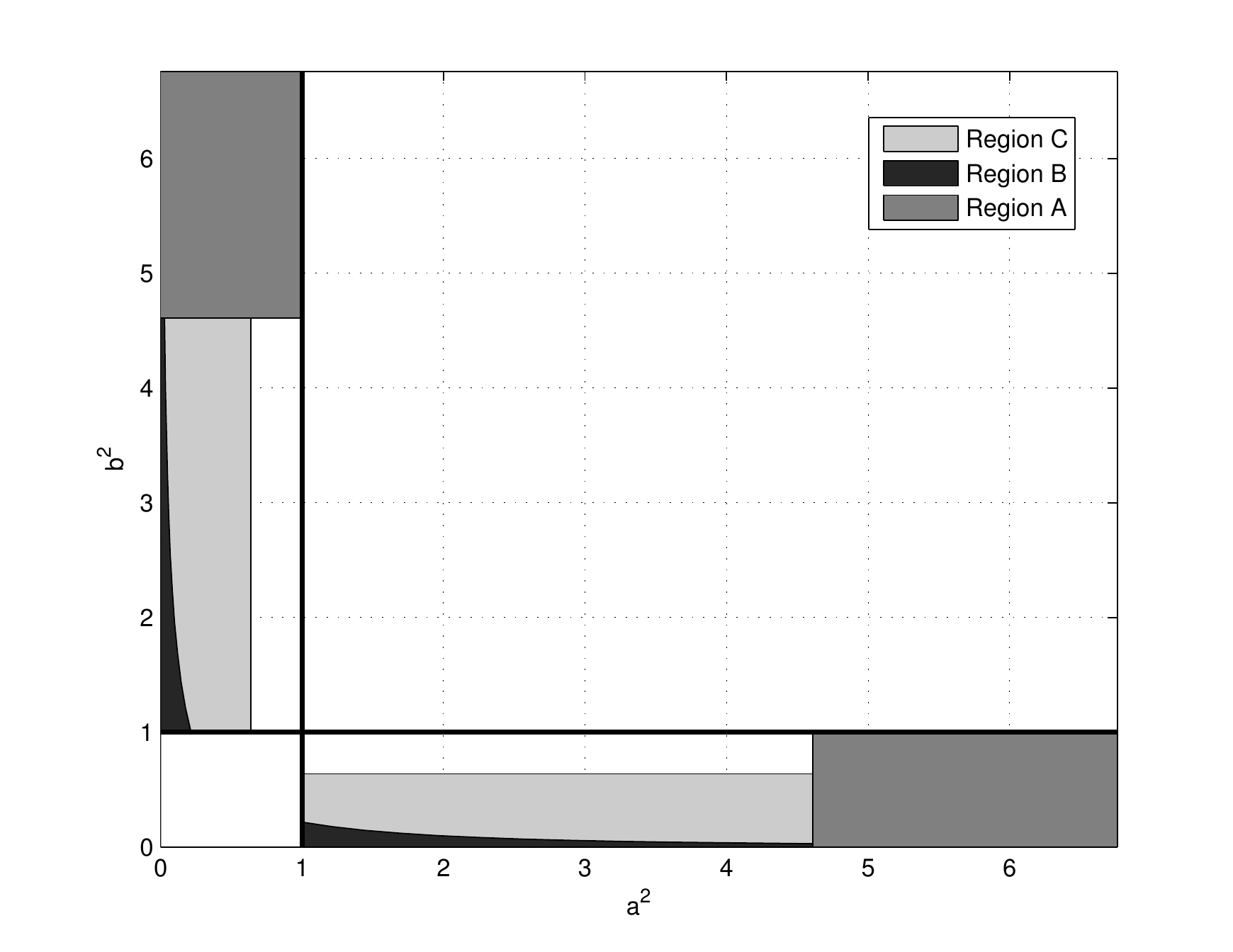}
\caption{Region where sum capacity is achieved within 0.2 bit, $P_1 = -3$ dB, $P_2 = -3$ dB.}
\label{fig:a2b2planehalf}
\end{figure}
 Fig. \ref{fig:a2b2planehalf} shows the region in the $(a^2,b^2)$ plane where MAC at receiver 1 or MAC at receiver 2 achieves within 0.2 bit of the sum capacity with power constraints $P_1=-3$ dB, $P_2=-3$ dB.  Region A corresponds to the regions obtained in Theorems \ref{thm1} and \ref{thm4}. Region B corresponds to the regions obtained in Theorems \ref{thm2} and \ref{thm5}.  Region C corresponds to the regions obtained in Theorems \ref{thm3} and \ref{thm6}. It can be seen that a significant part of the mixed interference regime is covered by these regions. 

\begin{figure}[htbp]
\centering
\includegraphics[width=0.95\columnwidth]{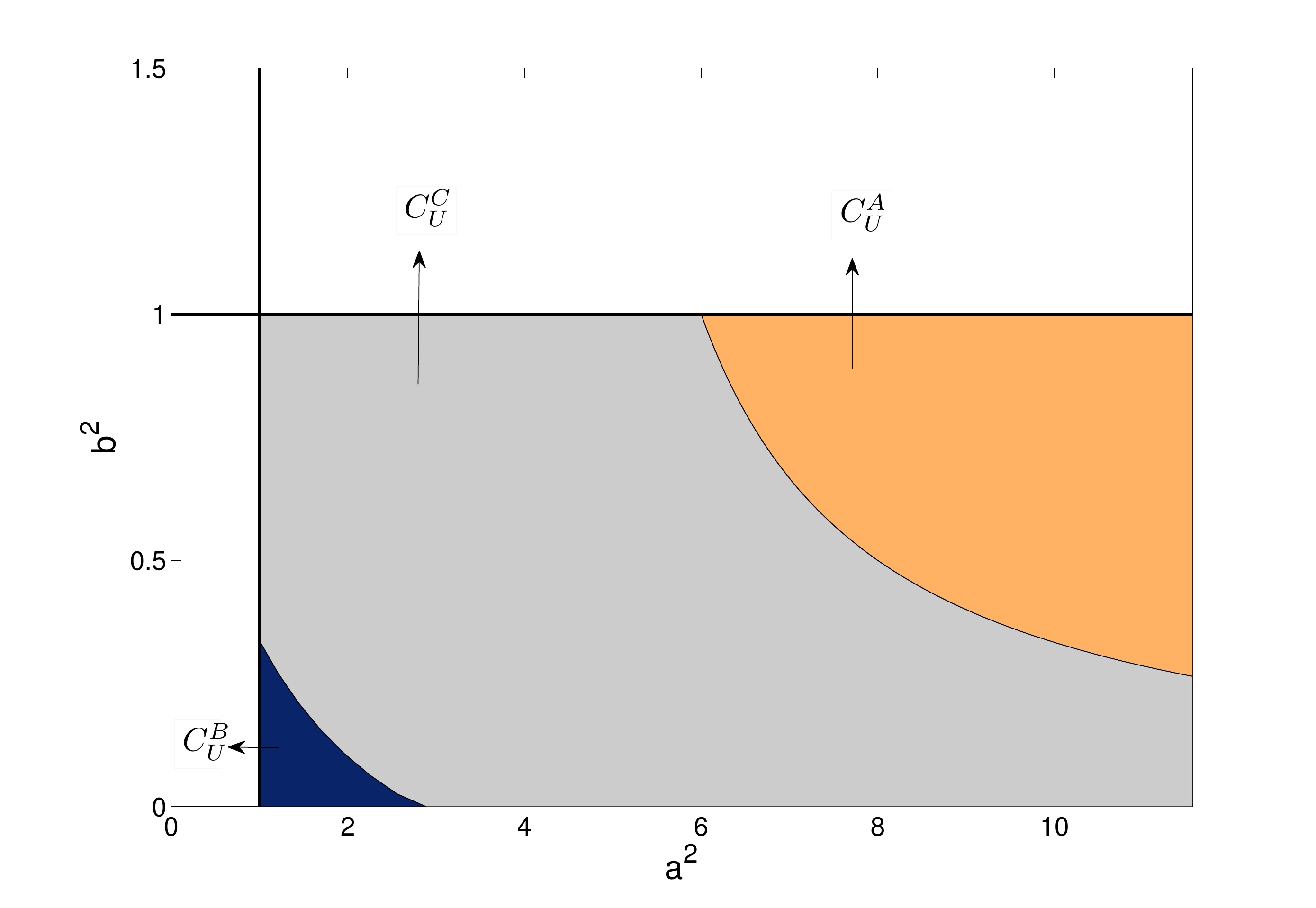}
\caption{Subregions of ${\cal R}_1$ where each bound is minimum: $P_1 = -0$ dB, $P_2 = -5$ dB}
\label{fig:regionsABC}
\end{figure}

Fig. \ref{fig:regionsABC} shows the subregions in ${\cal R}_1$ where each of the three sum capacity bounds is minimum for $P_1 = 0$ dB, $P_2 = -5$ dB.
\begin{figure}[htbp]
\centering
\includegraphics[width=0.95\columnwidth]{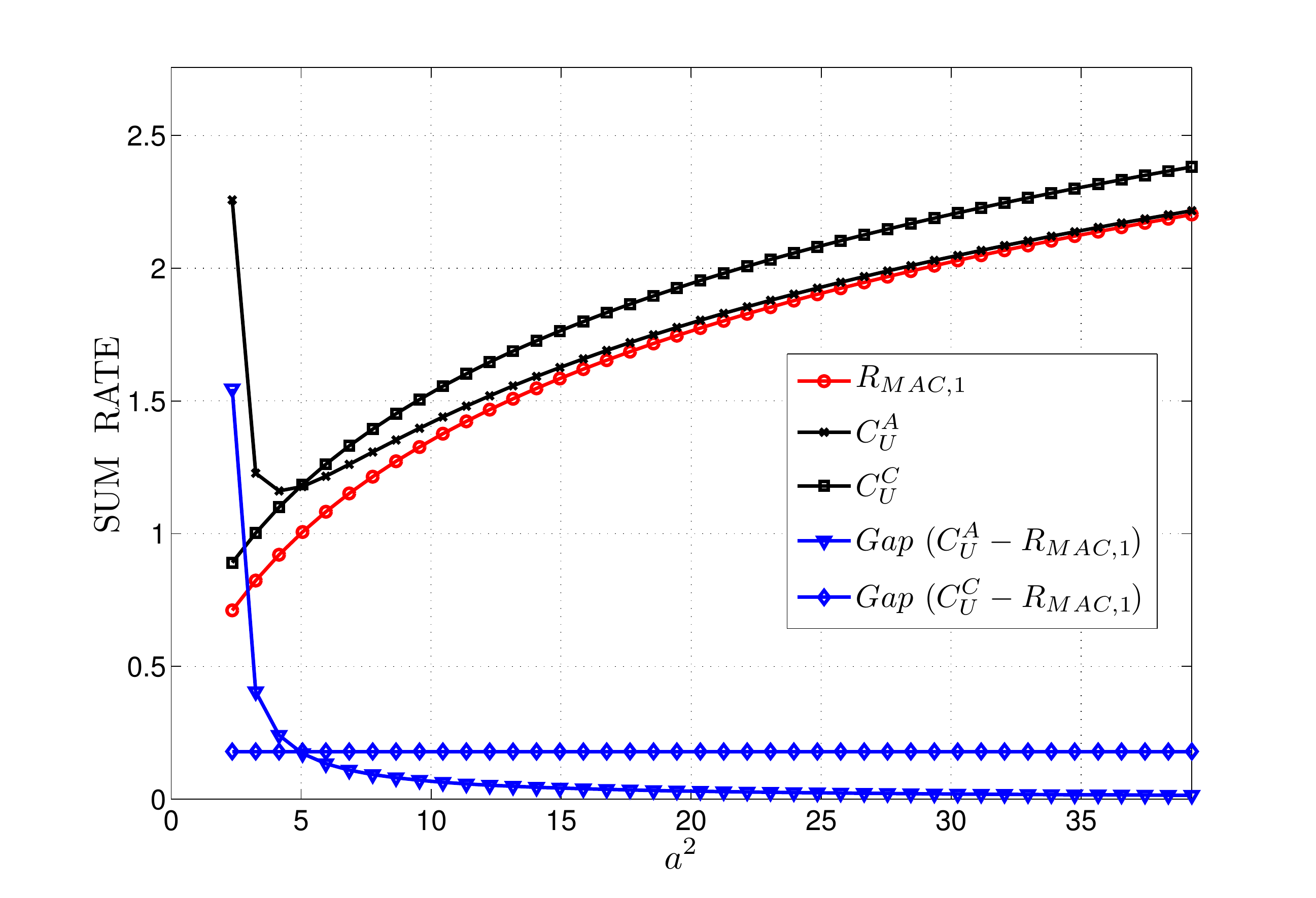}
\caption{Comparison of sum rate of MAC at receiver 1 and sum capacity upper bounds: $P_1 = -3$ dB, $P_2=-3$ dB, $b = 0.75$.}
\label{fig:sumratea2}
\end{figure}
Fig. \ref{fig:sumratea2} compares the sum rate achieved by MAC at receiver 1 with the upper bounds derived in Theorem \ref{thm1}(a) and \ref{thm3}(a) for different values of $a^2$. The gap is also plotted for each bound. The gap with $C_U^A$ reduces rapidly with increasing $a^2$. Note that $C_U^A$ does not depend on the value of $P_2$ and $b$, as long as $b^2 \le 1$. However, $C_U^C$ depends on $b$. 

\begin{figure}[htbp]
\centering
\includegraphics[width=0.9\columnwidth]{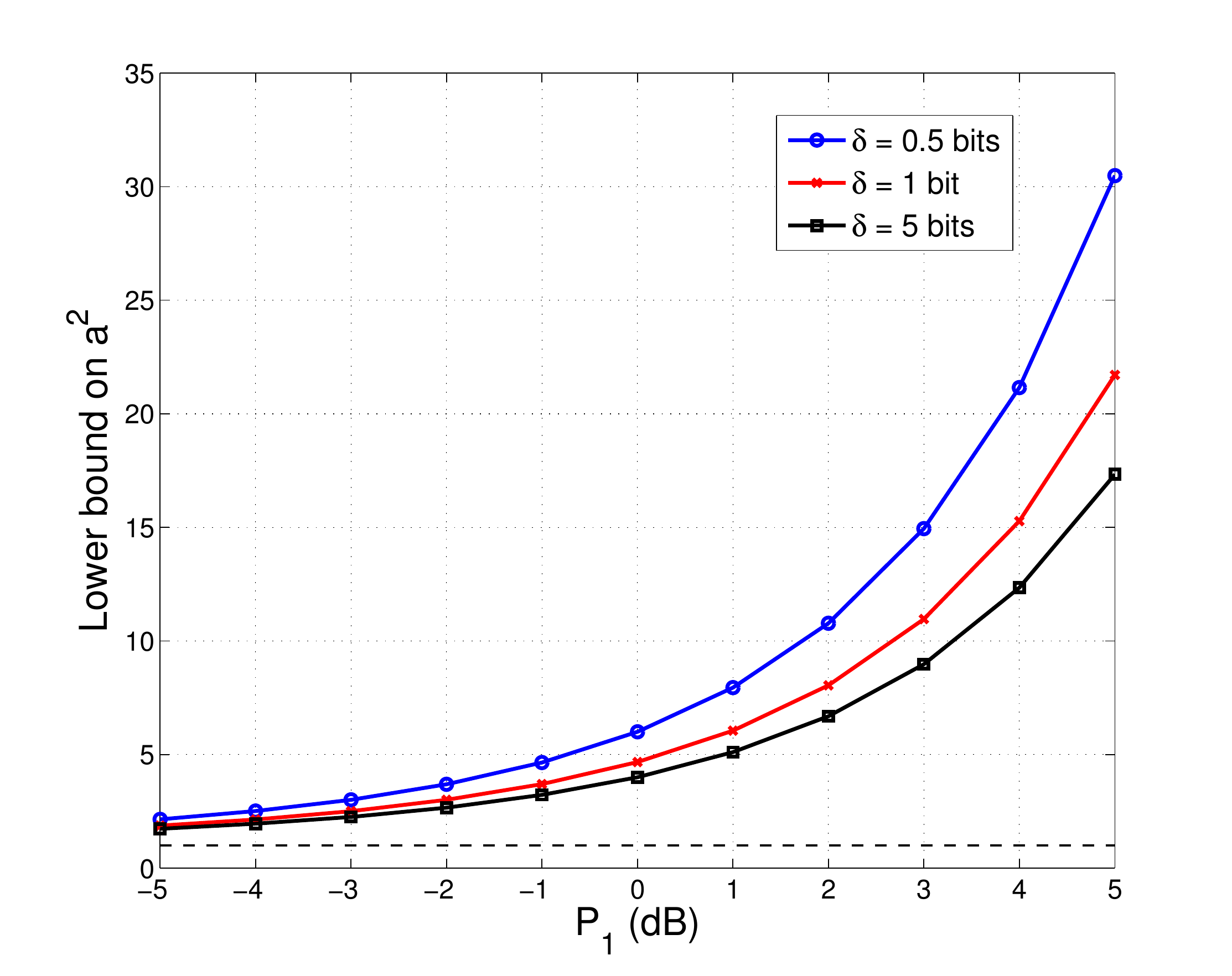}
\caption{Lower bound on $a^2$ in Theorem \ref{thm1}(b) as a function of $P_1$ for different values of $\delta$}
\label{fig:a2versusP1}
\end{figure}

Finally, in Fig. \ref{fig:a2versusP1}, the lower bound on $a^2$ in Theorem \ref{thm1}(b) is plotted as a function of $P_1$ for various values of $\delta$.

\section{Conclusions}
The sum capacity of the Gaussian X channel is exactly known only in the noisy interference regime \cite{HuaCadJaf12}. The approximate capacity of the Gaussian X channel within a gap of 4 bits was recently obtained in \cite{NieMad13}. In \cite{NieMad13}, the focus was on channel conditions where interference alignment is required. In this paper, we focused on the Gaussian X channel in the mixed interference regime and showed that multiple access transmission to one of the receivers is close to optimal in terms of sum rate in this regime. For any given $\delta > 0$, the subregion of the mixed interference regime where the achieved sum rate is within $\delta$ bits of sum capacity was determined. Three upper bounds were derived and the subregions where each bound is tighter than the other two were obtained.

\vspace*{-4mm}
\bibliographystyle{IEEEtran}
\bibliography{refsmixedx}

\end{document}